\documentclass[letterpaper,conference]{IEEEtran}
\usepackage{cite}
\usepackage{amsmath,amssymb,amsfonts}
\usepackage{algorithmic}
\usepackage{graphicx}
\usepackage{textcomp}
\usepackage[table]{xcolor}
\usepackage[caption=false]{subfig}
\usepackage{amsmath}
\usepackage{multirow}
\usepackage{soul}
\usepackage[]{nohyperref}  
\usepackage{url}  

\def\BibTeX{{\rm B\kern-.05em{\sc i\kern-.025em b}\kern-.08em
	T\kern-.1667em\lower.7ex\hbox{E}\kern-.125emX}}
\begin{document}
\title{Transforming Agriculture: Exploring Diverse Practices and Technological Innovations}
	\author{\IEEEauthorblockN{Ramakant Kumar}
	}
\maketitle
\begin{abstract}
Agriculture is a vital sector that significantly contributes to the economy and food security, particularly in regions like Varanasi, India. This paper explores various types of agriculture practiced in the area, including subsistence, commercial, intensive, extensive, industrial, organic, agroforestry, aquaculture, and urban agriculture. Each type presents unique challenges and opportunities, necessitating innovative approaches to enhance productivity and sustainability. To address these challenges, the integration of advanced technologies such as sensors and communication protocols is essential. Sensors can provide real-time data on soil health, moisture levels, and crop conditions, enabling farmers to make informed decisions. Communication technologies facilitate the seamless transfer of this data, allowing for timely interventions and optimized resource management. Moreover, programming techniques play a crucial role in developing applications that process and analyze agricultural data. By leveraging machine learning algorithms, farmers can gain insights into crop performance, predict yields, and implement precision agriculture practices. This paper highlights the significance of combining traditional agricultural practices with modern technologies to create a resilient agricultural ecosystem. The findings underscore the potential of integrating sensors, communication technologies, and programming in transforming agricultural practices in Varanasi. By fostering a data-driven approach, this research aims to contribute to sustainable farming, enhance food security, and improve the livelihoods of farmers in the region.
\end{abstract}
   \section{Introduction}

Agriculture plays a pivotal role in the socio-economic fabric of countries, particularly in developing regions like India. With over 58\% of the rural population engaged in agricultural activities, it is essential to explore various farming practices that can enhance productivity, sustainability, and food security. This paper examines the diverse types of agriculture prevalent in India, India, emphasizing the need for innovative solutions to address the challenges faced by farmers in the region~\cite{7377400,gupta2024utilizingtransferlearningpretrained,1321026}.

The agricultural landscape in India is characterized by a blend of traditional and modern practices. Subsistence agriculture remains prominent, where farmers primarily cultivate crops for family consumption, often utilizing age-old techniques that are increasingly challenged by climate variability and resource constraints. In contrast, commercial agriculture is gaining traction, driven by market demands and technological advancements. However, the effective integration of high-yield methods, mechanization, and data-driven practices is crucial for maximizing outputs and ensuring economic viability.

As the global agricultural sector evolves, the adoption of technology becomes imperative. Advanced communication protocols, sensors, and artificial intelligence (AI) have the potential to revolutionize agricultural practices~\cite{conf/networking/PandeyKGRR24,8480255,conf/wcnc/PandeyKG024}. For instance, precision agriculture leverages data analytics to optimize resource use, reduce waste, and improve crop yields. In India, the integration of IoT technologies and machine learning can facilitate real-time monitoring of soil health, weather conditions, and crop performance, enabling farmers to make informed decisions.

Furthermore, sustainable practices such as organic farming, agroforestry, and permaculture offer alternatives that prioritize ecological balance and biodiversity. These approaches not only contribute to environmental health but also enhance the resilience of farming systems against climate change impacts. By embracing a multi-faceted agricultural strategy that combines traditional knowledge with modern innovations, India can pave the way for a more sustainable and prosperous agricultural future.

This paper aims to explore the various types of agriculture in India, highlighting their unique characteristics, challenges, and the trans-formative role of technology. By understanding these dynamics, we can better inform agricultural policies and practices that support the livelihoods of farmers while promoting sustainability and food security in the region.
\section{Agriculture Types}
Agriculture is a diverse field encompassing various practices tailored to meet the needs of different communities, climates, and economic goals. Each type of agriculture employs unique techniques, technologies, and management strategies, which can significantly influence sustainability and productivity. Understanding these types is crucial for optimizing agricultural practices and implementing effective policies ~\cite{10176016,conf/wcnc/ThangadoraiSGK24,conf/wcnc/0001G24,5558084}.
\begin{figure}[h]
    \centering
    \includegraphics[width=1\linewidth]{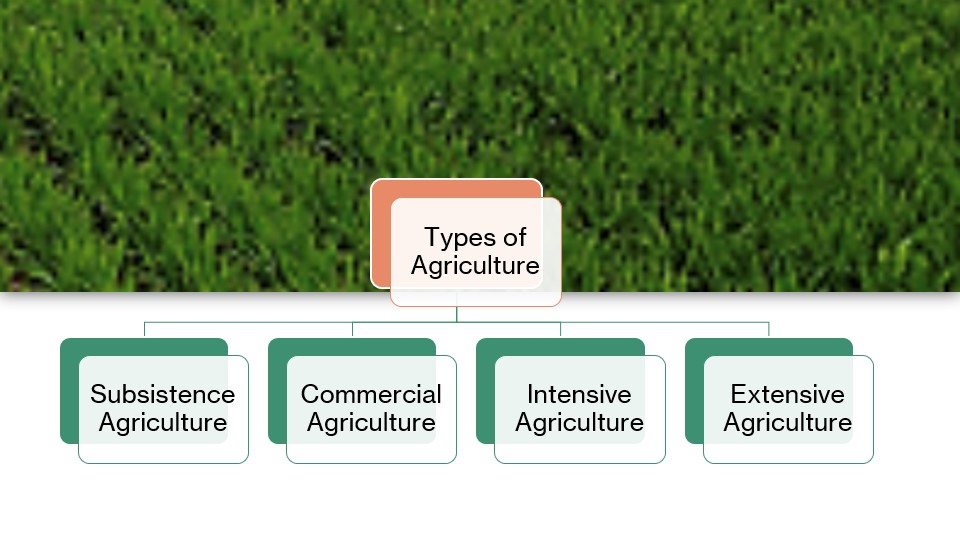}
    \caption{Types Of Agriculture}
\end{figure}

\begin{table*}[h]
    \centering
    \caption{Types of Agriculture and Their Characteristics}
    \begin{tabular}{|l|l|l|l|}
        \hline
        \textbf{Type of Agriculture} & \textbf{Characteristics} & \textbf{Technology Integration} & \textbf{Challenges} \\ \hline
        Subsistence Agriculture & Focused on family consumption, minimal surplus & Local sensors for soil health & Limited access to resources \\ \hline
        Commercial Agriculture & Grown for sale, high-yield methods & IoT for crop monitoring & Market fluctuations \\ \hline
        Intensive Agriculture & High input and labor, uses fertilizers & Precision agriculture tools & Resource management \\ \hline
        Extensive Agriculture & Low input per land area, natural fertility & Remote sensing for monitoring & Land availability \\ \hline
        Industrial Agriculture & Mass production, often monocropping & Automated systems and analytics & Environmental concerns \\ \hline
        Organic Agriculture & Avoids synthetic chemicals, sustainable practices & Organic farming technologies & Certification costs \\ \hline
        Agroforestry & Combines trees with crops/livestock & Ecological monitoring systems & Land-use competition \\ \hline
        Aquaculture & Cultivation of aquatic organisms & Water quality sensors & Disease management \\ \hline
        Permaculture & Imitates natural ecosystems, sustainable & Design software for planning & Initial setup complexity \\ \hline
        Horticulture & Focused on fruits, vegetables & Smart irrigation systems & Seasonal variability \\ \hline
        Precision Agriculture & Data-driven farming techniques & GPS, IoT sensors & High setup costs \\ \hline
        Mixed Farming & Crop and livestock integration & Data analytics for synergies & Knowledge transfer \\ \hline
        Urban Agriculture & Food production in urban settings & Vertical farming tech & Space limitations \\ \hline
        Dryland Farming & Drought-resistant crops, water conservation & Soil moisture sensors & Weather variability \\ \hline
        Shifting Cultivation & Rotating cultivation areas & GIS for planning & Soil nutrient depletion \\ \hline
    \end{tabular}
    \label{tab:agriculture_types}
\end{table*}
\subsection{Subsistence Agriculture} 
Farming primarily for the farmer's family consumption, with little to no surplus for sale. Common in developing regions and often involves traditional methods. 

Subsistence agriculture in India can benefit from the integration of modern technology to enhance productivity and sustainability. Sensors for soil moisture, temperature, and nutrient levels are increasingly being used to provide farmers with real-time data on their fields. These sensors help optimize water usage and improve soil health, which is essential for regions where water resources may be limited.

Furthermore, low-cost communication technologies, such as LoRaWAN, enable data collected from these sensors to be transmitted over long distances with minimal energy consumption. This connectivity is especially beneficial in rural areas with limited internet infrastructure, allowing small-scale farmers to access information on crop health, weather forecasts, and soil conditions through smartphone applications. By adopting these technologies, subsistence farmers in India can improve crop yields while minimizing resource wastage, thereby enhancing the sustainability of their practices~\cite{Qin2020,Shubyn2023,1589116,journals/tpds/MishraGBD24,7815384,conf/sensys/KumariG023,7488250}.

\subsection{Commercial Agriculture} 
Focused on growing crops and raising livestock for sale in markets, commercial agriculture in India includes high-value crops like rice, wheat, and sugarcane. These practices benefit from advanced technology to enhance productivity, efficiency, and profitability.

In India, commercial agriculture leverages IoT-based sensors to monitor soil pH, moisture, temperature, and nutrient levels, allowing for precision farming that optimizes input use. LoRaWAN and NB-IoT networks connect these sensors over large fields, enabling real-time data collection and remote monitoring, which is crucial for managing extensive farm sizes. Machine learning algorithms analyze this sensor data to forecast crop yields, identify early signs of pests or diseases, and optimize irrigation and fertilization schedules. For livestock farming, AI-driven applications monitor animal health and optimize feeding, improving productivity and reducing waste~\cite{conf/iccps/0012G023,9166711}.

Furthermore, farmers use mobile applications connected to cloud platforms to receive timely insights into market prices, weather patterns, and recommendations on resource management. By integrating these intelligent solutions, commercial farms can enhance yield quality, reduce input costs, and respond proactively to environmental changes, thus boosting profitability and sustainability in  India.

\subsection{Intensive Agriculture} 
Characterized by high input and labor per land area, intensive agriculture in India often includes practices such as greenhouse agriculture, multi-cropping, and precision farming, aimed at maximizing output from limited space.

In India, IoT sensors are used extensively in greenhouses and crop fields to monitor critical parameters like temperature, humidity, soil moisture, and nutrient levels. These sensors connect through LoRaWAN or Wi-Fi networks to allow continuous data collection across intensive agricultural sites. Machine learning models analyze this data in real time to automate irrigation, ventilation, and nutrient delivery systems, which helps maintain optimal growing conditions for high-yield crops.

AI-based applications also enable pest and disease detection by analyzing visual and sensor data from crops, providing early warnings to farmers. Additionally, predictive analytics powered by machine learning optimize crop cycles and recommend best practices based on historical data, soil type, and seasonal patterns. Through mobile applications, farmers can manage these systems remotely, adjusting settings as needed to ensure consistent and high-quality yields, thereby enhancing productivity and resource efficiency in intensive agricultural practices in India~\cite{conf/icc/KumariGDB23,journals/csur/MishraG23,1589116,9311219,9479778,journals/wpc/ChopadeGD23}.

\subsection{Extensive Agriculture} 
Extensive agriculture relies on large tracts of land with minimal input per unit area, often using natural soil fertility and low labor input. In India, extensive agriculture is commonly practiced in areas with abundant land and low population density, often for crops like rice, wheat, and sugarcane, as well as livestock grazing.

To support extensive agriculture in India, GPS-enabled IoT sensors measure soil health indicators like moisture, pH, and nutrient content across large fields. These sensors connect via LoRaWAN or GSM networks, allowing data to be collected remotely and transmitted to central systems, where AI algorithms analyze spatial trends in soil quality and crop health. AI-based recommendations inform farmers of areas needing specific inputs, such as additional fertilizer or water, and help prevent resource overuse by targeting interventions only where necessary~\cite{conf/infocom/MishraGD22,Wang2021}.

Additionally, machine learning models predict pest outbreaks, weather changes, and crop yield estimates based on historical and current sensor data, enabling farmers to make proactive management decisions. Livestock in extensive grazing systems also benefit from GPS-based monitoring, where AI-powered applications can recommend grazing patterns that prevent overgrazing and optimize land use. With these technologies, extensive agriculture in India can become more sustainable and resource-efficient, balancing productivity with conservation of land and resources.

\subsection{Industrial Agriculture} 
Large-scale, technology-intensive farming aimed at mass production, often involving monocropping and confined animal feeding operations. In India, industrial agriculture includes large-scale crops like rice and wheat, with advanced technology integrated to optimize productivity and efficiency.

IoT sensors and remote sensing devices monitor soil health, crop growth, and environmental conditions, enabling precise control over large fields. These sensors are linked via LoRaWAN and cellular networks, allowing data to be transmitted continuously to centralized platforms for analysis. AI algorithms process this data to optimize planting, irrigation, and fertilization schedules, improving yield and reducing resource wastage. This data-driven approach minimizes operational costs by ensuring that inputs are used only as needed, leading to substantial savings for farmers.

Moreover, by enhancing crop resilience and productivity, industrial agriculture can lower the risk of crop failure, contributing to long-term sustainability. However, challenges such as soil degradation from monocropping practices remain. Sustainable practices, such as crop rotation and integrated pest management, must be incorporated to maintain soil health. With these technology-driven methods, industrial agriculture in India becomes more scalable, efficient, and productive, catering to both local and national demand while addressing sustainability concerns~\cite{conf/wowmom/KumariGDD22}.

\subsection{Organic Agriculture} 
Avoids synthetic chemicals, focusing on natural fertilizers and pest control methods, prioritizing sustainability and ecological balance. In India, organic farming practices are supported by data-driven tools to ensure ecological integrity and crop health.

IoT sensors monitor soil health, focusing on organic parameters like moisture, temperature, and pH, which are essential for maintaining a healthy, chemical-free growing environment. Connected via LoRaWAN or GSM networks, these sensors provide real-time data to farmers, who can make adjustments using AI-based insights to enhance soil fertility naturally. While the initial investment in organic farming may be higher due to the costs of natural inputs and technology adoption, long-term benefits include reduced chemical input costs and increased market prices for organic produce~\cite{conf/mswim/KumariGM021}.

Machine learning algorithms analyze historical data to recommend crop rotations and companion planting that support pest resistance without synthetic inputs. This not only enhances biodiversity but also promotes soil health and reduces reliance on external inputs, leading to a more sustainable farming system. AI-powered mobile applications help farmers track crop health and monitor for pest and disease presence, offering eco-friendly treatment suggestions. By promoting sustainability and reducing environmental impact, organic agriculture in India can improve yield quality while preserving biodiversity and soil health~\cite{gupta2024lessonslearnedsmartcampus}.

\subsection{Agroforestry} 
Integrates trees and shrubs with crops or livestock, enhancing biodiversity and sustainability while providing diverse outputs like food, fuel, and fodder. In India, agroforestry systems incorporate native trees and crops, which benefit from integrated technology for sustainability and productivity.

IoT sensors track soil moisture, nutrient levels, and microclimate conditions across different tree-crop zones. These sensors use LoRaWAN or Wi-Fi networks to send data to central monitoring platforms, where AI-based systems analyze the micro-ecosystem dynamics, supporting decisions on water and nutrient allocation. Machine learning models suggest optimal intercropping strategies based on historical and real-time data, increasing crop resilience and improving resource efficiency~\cite{10176016,conf/wowmom/MishraKGSDSP21,4460126,conf/infocom/KumariGD21}.

In terms of cost, agroforestry can reduce overall agricultural expenses by diversifying income sources and minimizing reliance on chemical inputs. Sustainable practices like tree planting enhance soil fertility, improve water retention, and provide additional products such as fruits and timber, creating a more resilient farming system. Additionally, agroforestry contributes to carbon sequestration and enhances biodiversity, making it a crucial strategy for sustainable land use in India. By integrating advanced technologies, agroforestry can promote both ecological sustainability and economic viability for farmers in the region~\cite{conf/icc/KumariGD20,journals/cem/GhoshMGSR22}.

\subsection{Aquaculture} 
The cultivation of aquatic organisms like fish, shellfish, and algae, used for food, medicinal products, and other resources. In India, aquaculture is gaining traction as a viable solution to enhance food security and provide alternative sources of income for local farmers.

Modern aquaculture systems utilize IoT-based sensors to monitor water quality parameters such as temperature, pH, dissolved oxygen, and nutrient levels. These sensors are interconnected through LoRaWAN or GSM networks, allowing for real-time monitoring of aquatic environments. AI-driven analytics interpret this data, providing farmers with actionable insights to optimize feeding schedules, manage water quality, and prevent disease outbreaks. For instance, machine learning algorithms can predict optimal feeding times and amounts based on growth patterns and environmental conditions, reducing feed costs and enhancing growth rates~\cite{9166711,9130098}.

Sustainability is a key focus in India's aquaculture practices, where integrated systems such as aquaponics—combining fish farming with hydroponics—are being explored. This method not only produces fish but also grows plants in nutrient-rich water, minimizing waste and promoting resource efficiency. Additionally, community-based aquaculture projects encourage local participation and knowledge sharing, further enhancing the sustainability of these systems. By reducing the dependency on wild fish stocks and increasing local production, aquaculture can significantly contribute to food security and economic stability in the region~\cite{9479778,9164991}.

\subsection{Permaculture} 
A system that imitates natural ecosystems, emphasizing sustainability, biodiversity, and self-sufficiency. In India, permaculture practices focus on creating sustainable agricultural systems that work in harmony with the environment, utilizing natural resources efficiently.

Permaculture designs in India often incorporate polycultures, where multiple species of plants and animals coexist to support one another, leading to enhanced biodiversity and resilience against pests and diseases. IoT sensors can be used to monitor soil health, moisture levels, and microclimate conditions, providing data that informs decisions on planting and crop management. These sensors, connected through robust communication networks, allow farmers to track changes in the ecosystem in real-time, promoting proactive management strategies~\cite{10122600,1589116}.

AI and machine learning models analyze data from these sensors to optimize resource allocation, such as water and nutrients, based on the specific needs of different plant and animal species. This data-driven approach helps maintain a balance within the ecosystem, ensuring sustainable yields while minimizing input costs.

Permaculture in India also emphasizes water conservation techniques, such as rainwater harvesting and swales, which enhance soil moisture retention and reduce irrigation needs. By implementing these practices, farmers can create self-sustaining systems that provide food and resources while improving environmental health. The integration of technology in permaculture promotes a more efficient use of resources, leading to long-term sustainability and resilience against climate change~\cite{9311219,4460126}.

\subsection{Horticulture} 
Focused on growing fruits, vegetables, flowers, and ornamental plants, often on a smaller scale and with intensive methods. In India, horticulture plays a vital role in local economies and contributes to food security and nutritional diversity.

Advancements in technology, such as IoT sensors and smart irrigation systems, are transforming horticultural practices in the region. Sensors monitor soil moisture, temperature, and nutrient levels, allowing for precision irrigation that optimizes water usage and reduces costs. These sensors communicate via LoRaWAN networks, enabling farmers to make informed decisions based on real-time data. AI algorithms analyze this data to forecast yield potentials, identify optimal harvest times, and recommend crop rotation strategies, enhancing both productivity and sustainability~\cite{7745306,8480255}.

In addition, machine learning applications help identify diseases and pests through image recognition technology, allowing for timely interventions that minimize chemical use and enhance crop health. The integration of sustainable practices, such as organic fertilizers and integrated pest management, aligns with the growing demand for eco-friendly products in local and national markets.

Horticulture in India also benefits from community-supported agriculture (CSA) models, where local consumers subscribe to receive fresh produce directly from farmers. This not only supports local economies but also fosters a connection between consumers and producers, promoting sustainable practices. By incorporating technology and sustainable practices, horticulture in India can improve yield quality, reduce costs, and enhance environmental health, contributing to a more resilient agricultural landscape~\cite{8170296,8502812}.

\subsection{Precision Agriculture} 
Uses technology like GPS, IoT sensors, and data analytics to optimize crop yields, reduce waste, and improve resource use efficiency. In India, precision agriculture is revolutionizing farming practices, allowing local farmers to maximize productivity while minimizing environmental impacts.

IoT sensors deployed across fields collect real-time data on soil moisture, nutrient levels, and crop health. These sensors utilize LoRaWAN or GSM networks for data transmission, ensuring constant monitoring and timely decision-making. GPS technology helps in creating detailed maps of fields, enabling farmers to identify variability in soil conditions and crop performance. This information is processed using AI algorithms that recommend targeted interventions, such as variable rate irrigation and fertilization, ensuring that resources are applied precisely where needed~\cite{7880946,10176016}.

Additionally, drone technology is increasingly utilized in precision agriculture to monitor crop health and assess field conditions from the air. Drones equipped with multispectral cameras can identify areas of stress or disease early, allowing for prompt action to be taken. By implementing these advanced technologies, farmers in India can significantly increase their crop yields while reducing water and fertilizer usage, leading to more sustainable agricultural practices~\cite{7815384,7803607}.

\subsection{Mixed Farming} 
Combines crop cultivation and livestock rearing on the same land, creating synergies between the two. In India, mixed farming systems are gaining popularity due to their ability to enhance farm resilience and sustainability.

By integrating crops and livestock, farmers can utilize animal manure as a natural fertilizer, reducing dependency on chemical inputs. IoT sensors can monitor soil health and nutrient levels, helping farmers optimize the application of manure and other organic fertilizers. Additionally, sensors can track the health and productivity of livestock, providing data that informs feeding strategies and breeding decisions.

This system allows for the efficient use of resources, as crop residues can be fed to livestock, while livestock can graze on cover crops, improving soil health and preventing erosion. Machine learning models analyze data from mixed farming operations to optimize planting schedules and livestock management, enhancing overall farm productivity. The combined approach also helps farmers diversify their income streams, making them more resilient to market fluctuations and climatic challenges~\cite{7377400,7123563,7488250}.

\subsection{Urban Agriculture} 
Involves growing food in urban areas, often through rooftop gardens, hydroponics, and vertical farming to meet urban demand locally. In India, urban agriculture is emerging as a critical solution to food security in densely populated areas.

Rooftop gardens and vertical farms utilize limited space efficiently, producing fresh vegetables and herbs for local communities. IoT technology plays a crucial role in these systems, where sensors monitor environmental conditions such as light, humidity, and temperature to optimize growing conditions. Hydroponic systems use nutrient-rich water solutions instead of soil, requiring significantly less water compared to traditional farming methods.

AI-driven analytics assist in managing these urban farms by predicting optimal planting times and monitoring plant health. Community engagement is essential in urban agriculture, with local residents often participating in gardening initiatives that foster social connections and promote awareness of sustainable practices. By implementing these technologies and practices, urban agriculture in India not only addresses food scarcity but also enhances community resilience and promotes sustainability in urban environments~\cite{7498684,8016573}.

\subsection{Dryland Farming} 
Specialized in areas with low rainfall, focusing on drought-resistant crops and water conservation techniques. In India, where variability in rainfall patterns can significantly impact agricultural productivity, dryland farming practices are becoming increasingly relevant.

Farmers in the region employ drought-resistant crop varieties, such as millets and legumes, which are well-suited to arid conditions. Techniques like rainwater harvesting and contour farming are implemented to maximize water retention in the soil. IoT sensors monitor soil moisture levels, enabling farmers to make informed irrigation decisions and optimize water usage.

AI algorithms analyze historical climate data to predict rainfall patterns, helping farmers plan their planting and harvesting schedules accordingly. These data-driven insights support farmers in implementing efficient water conservation practices, ensuring sustainable use of available resources. By adopting these methods, dryland farming in India not only enhances food security but also contributes to the sustainable management of water resources in a changing climate~\cite{6054047,5558084}.

\subsection{Shifting Cultivation} 
Involves rotating cultivation areas to allow soil to regain nutrients, common in forested and tropical areas. In India, shifting cultivation is practiced in some tribal and rural communities, emphasizing traditional knowledge and sustainable land management.

Farmers rotate their fields, allowing fallow periods that help restore soil fertility naturally. During these fallow periods, IoT sensors can monitor soil health and biodiversity, providing data that informs future cultivation decisions. This practice helps prevent soil degradation and enhances the resilience of farming systems.

Integrating modern techniques with traditional practices, machine learning models can analyze data on soil fertility and crop performance over time, guiding farmers on the best rotation schedules and crop combinations. By promoting biodiversity and maintaining ecological balance, shifting cultivation in India can contribute to sustainable agriculture while respecting local traditions and practices~\cite{8326735,7460727,Wang2021}.

\section{Types of Communication Protocols for Agriculture}

Communication protocols play a vital role in modern agriculture, enabling seamless data exchange among various devices and systems. The primary types of communication protocols used in agriculture include:
\begin{figure}[h]
    \centering
    \includegraphics[width=1\linewidth]{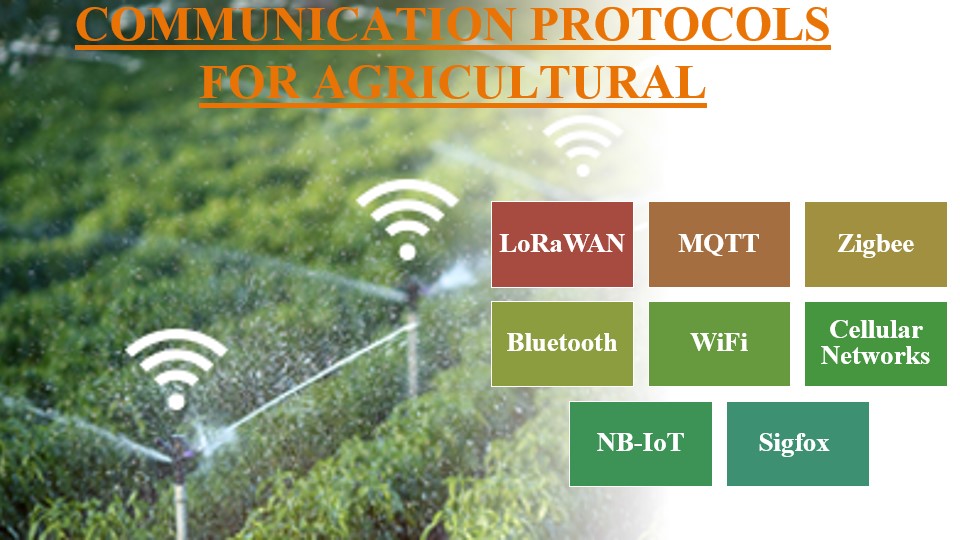}
    \caption{TYPES OF COMMUNICATION PROTOCOLS}
    \label{fig:enter-label}
\end{figure}
\begin{table*}[h]
    \centering
    \caption{Communication Protocols for Agricultural Applications}
    \begin{tabular}{|p{2cm}|p{4.5cm}|p{4.5cm}|p{4.5cm}|}
        \hline
        \textbf{Protocol} & \textbf{Characteristics} & \textbf{Applications} & \textbf{Challenges} \\ \hline
        LoRaWAN & Low-power, wide-area network suitable for IoT & Connecting sensors over long distances & Limited data rate \\ \hline
        MQTT & Lightweight messaging protocol for low-bandwidth & Communication between IoT devices and cloud services & Requires reliable network connectivity \\ \hline
        Zigbee & Short-range, low-power wireless communication & Sensor networks and home automation & Interference from other devices \\ \hline
        Bluetooth & Short-range communication, low energy consumption & Connecting mobile applications to sensors & Range limitations \\ \hline
        Wi-Fi & High-speed internet access and data transfer & Precision agriculture applications & Signal range and bandwidth issues \\ \hline
        Cellular Networks & Wide coverage, high-speed communication & Remote monitoring and control of farm operations & Data costs and signal reliability \\ \hline
        NB-IoT & Narrowband IoT technology for long-range & Smart agriculture devices, soil monitoring & Deployment costs and infrastructure requirements \\ \hline
        Sigfox & Ultra-narrowband technology for IoT & Low-power applications like livestock tracking & Limited bi-directional communication \\ \hline
    \end{tabular}
    \label{tab:communication_protocols}
\end{table*}

\subsection{LoRaWAN (Long Range Wide Area Network)}
LoRaWAN is a low-power, wide-area network protocol specifically designed for Internet of Things (IoT) applications, making it particularly suitable for connecting sensors and devices over long distances in agricultural settings. With the capability to transmit data over several kilometers—up to 15-20 km in rural areas—LoRaWAN enables extensive monitoring across large agricultural fields. Its low power consumption allows devices to operate on battery power for several years, significantly reducing the need for frequent battery replacements and ensuring long-term deployment of sensors. This scalability supports the simultaneous connection of numerous devices within a single network, facilitating the integration of various sensors that monitor parameters such as soil moisture, temperature, and crop health. The protocol employs end-to-end encryption to ensure data security and integrity during transmission, which is crucial for protecting sensitive agricultural data. Utilizing a star network architecture, where multiple end devices communicate with a central gateway, simplifies the network design and management. Overall, LoRaWAN provides real-time monitoring capabilities, cost-effectiveness, and remote accessibility, empowering farmers to make informed decisions and optimize their agricultural practices effectively~\cite{6780609,Charef2023}.

\subsection{MQTT (Message Queuing Telemetry Transport)}
MQTT is a lightweight messaging protocol optimized for low-bandwidth and high-latency networks, making it an ideal choice for facilitating communication between Internet of Things (IoT) devices and cloud services in agricultural environments. Designed for simplicity and efficiency, MQTT operates on a publish-subscribe model, where devices (or clients) can publish messages to a specific topic and subscribe to receive messages on topics of interest. This architecture allows for efficient use of network resources, reducing data traffic and minimizing power consumption—an essential feature for battery-operated devices commonly used in agriculture. MQTT supports Quality of Service (QoS) levels, enabling reliable message delivery even in challenging network conditions. In agricultural applications, MQTT can be used for real-time monitoring and control of various parameters, such as irrigation systems, soil moisture levels, and environmental conditions. Its lightweight nature ensures that even devices with limited processing power can easily implement the protocol, fostering interoperability among different IoT devices. Moreover, MQTT's ability to function seamlessly over unreliable networks enhances its utility in remote agricultural areas, enabling farmers to gather and analyze data effectively to optimize their operations, improve yields, and make informed decisions~\cite{Sikder2023,Samariya2023}.

\subsection{Zigbee}

Zigbee is a wireless communication protocol designed for short-range, low-power applications, making it particularly suitable for sensor networks within agricultural settings. Operating in the IEEE 802.15.4 standard, Zigbee enables devices to communicate wirelessly over distances of up to 100 meters indoors and 300 meters outdoors, depending on the environment and obstacles. Its low power consumption allows devices to function for extended periods on battery power, making it ideal for agricultural sensors that require long-term deployment without frequent maintenance. Zigbee utilizes a mesh network topology, where devices can communicate with one another and relay messages, extending the overall range and reliability of the network. This feature is especially beneficial in large agricultural fields where sensors may be deployed over considerable distances. 

In agriculture, Zigbee is commonly used for monitoring environmental conditions such as soil moisture, temperature, and humidity, enabling farmers to make data-driven decisions for irrigation and crop management. The protocol supports a large number of devices—up to 65,000 in a single network—allowing for comprehensive monitoring across diverse agricultural parameters. Furthermore, Zigbee’s interoperability with various devices facilitates the integration of sensors, actuators, and controllers into a cohesive network, streamlining farm management processes. The low latency of Zigbee communication ensures real-time data transmission, which is crucial for timely responses to changing environmental conditions. Overall, Zigbee's characteristics make it an essential tool for enhancing efficiency, productivity, and sustainability in modern agricultural practices~\cite{Kea2023,Rodriguez2023}.

\subsection{Bluetooth and BLE (Bluetooth Low Energy)}

Bluetooth and its low-energy variant, Bluetooth Low Energy (BLE), are wireless communication protocols designed for short-range connectivity, typically within a range of 10 to 100 meters. BLE is particularly advantageous for agricultural applications due to its low power consumption, allowing devices to operate for extended periods on small batteries. This makes it suitable for connecting mobile applications to various agricultural sensors and wearable devices, such as soil moisture sensors, temperature sensors, and livestock tracking systems. BLE’s efficient communication protocol allows for quick pairing and data transfer, enabling real-time monitoring of agricultural parameters. 

In agriculture, BLE can be used for applications such as remote monitoring of irrigation systems, tracking the health and location of livestock, and collecting data from various environmental sensors. Its ability to support multiple devices simultaneously enhances the flexibility of IoT deployments on farms. Moreover, BLE can connect to smartphones and tablets, allowing farmers to access and analyze data on-the-go, facilitating informed decision-making and timely interventions. With the growing trend of precision agriculture, the integration of Bluetooth technology into agricultural practices fosters enhanced data collection, improved resource management, and increased operational efficiency, ultimately contributing to more sustainable farming practices~\cite{Qin2020,Shubyn2023}.

\subsection{Wi-Fi}

Wi-Fi is a widely used wireless networking technology that enables high-speed internet access and data transfer, making it a crucial component of modern agricultural practices, particularly in precision agriculture applications. Wi-Fi operates in the 2.4 GHz and 5 GHz frequency bands, offering a range of coverage that can extend from a few meters to several hundred meters, depending on the network configuration and obstacles. Its ability to handle high data rates makes it ideal for transmitting large volumes of data from agricultural sensors, cameras, and drones in real-time. 

In agricultural settings, Wi-Fi facilitates the connection of various IoT devices, allowing farmers to monitor crop health, soil conditions, and environmental parameters effectively. By integrating Wi-Fi networks, farmers can utilize advanced technologies such as remote sensing, automated irrigation systems, and precision planting equipment that require reliable, high-speed internet connectivity. Additionally, Wi-Fi enables the use of cloud-based applications for data analysis, which enhances decision-making processes related to crop management and resource allocation. While Wi-Fi networks may require more power and infrastructure compared to other wireless protocols, their capability to support multiple high-bandwidth applications simultaneously makes them indispensable for modern agricultural operations that seek to optimize productivity and sustainability~\cite{5567086,1321026,Mohammadi2023}.

\section{Program for Agricultural Data Collection}
\begin{figure}[h]
    \centering
    \includegraphics[width=1\linewidth]{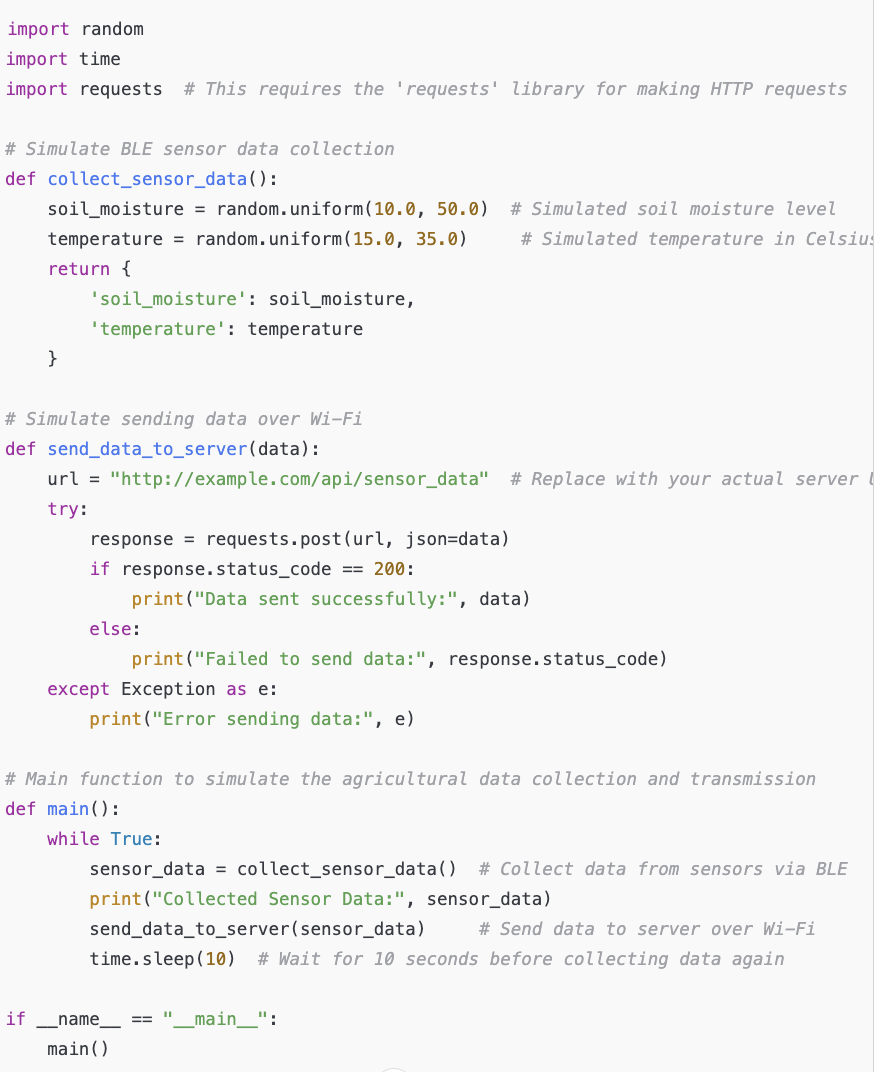}
    \caption{Program for Agricultural Data Collection}
\end{figure}
\subsection*{1. Importing Required Libraries}
The program starts by importing three libraries:
\begin{itemize}
    \item \texttt{random}: This library is used to generate random numbers, simulating sensor readings.
    \item \texttt{time}: This library is used to introduce delays in the program's execution, simulating periodic data collection.
    \item \texttt{requests}: This library is used to send HTTP requests to a server, allowing the program to transmit sensor data.
\end{itemize}

\subsection*{2. Function: \texttt{collect\_sensor\_data()}}
This function simulates the collection of sensor data:
\begin{itemize}
    \item It generates a random soil moisture level between 10.0 and 50.0.
    \item It generates a random temperature between 15.0 and 35.0.
    \item The function returns a dictionary containing the simulated values for soil moisture and temperature.
\end{itemize}

\subsection*{3. Function: \texttt{send\_data\_to\_server(data)}}
This function handles the transmission of sensor data to a server:
\begin{itemize}
    \item It defines the \texttt{url} variable, which should be replaced with the actual server endpoint for data collection.
    \item The function attempts to send the data using a POST request in JSON format.
    \item If the server responds with a status code of 200, it indicates successful transmission, and the data is printed.
    \item If the transmission fails, an error message with the status code is displayed.
    \item Any exceptions that occur during the request are caught and logged.
\end{itemize}

\subsection*{4. Function: \texttt{main()}}
This is the main function that runs the program:
\begin{itemize}
    \item It enters an infinite loop to continuously collect and send data.
    \item It calls \texttt{collect\_sensor\_data()} to obtain simulated sensor readings.
    \item It prints the collected sensor data to the console.
    \item It then calls \texttt{send\_data\_to\_server(sensor\_data)} to send the collected data to the server.
    \item The program pauses for 10 seconds before repeating the process, allowing for periodic data collection.
\end{itemize}

\subsection*{5. Running the Program}
The program is executed by calling the \texttt{main()} function within a \texttt{if \_\_name\_\_ == "\_\_main\_\_":} block, which ensures that the main function runs only when the script is executed directly, not when imported as a module.

\bibliographystyle{IEEEtran}
\bibliography{paper.bib}

\begin{thebibliography}{10}
\providecommand{\url}[1]{#1}
\csname url@samestyle\endcsname
\providecommand{\newblock}{\relax}
\providecommand{\bibinfo}[2]{#2}
\providecommand{\BIBentrySTDinterwordspacing}{\spaceskip=0pt\relax}
\providecommand{\BIBentryALTinterwordstretchfactor}{4}
\providecommand{\BIBentryALTinterwordspacing}{\spaceskip=\fontdimen2\font plus
\BIBentryALTinterwordstretchfactor\fontdimen3\font minus \fontdimen4\font\relax}
\providecommand{\BIBforeignlanguage}[2]{{%
\expandafter\ifx\csname l@#1\endcsname\relax
\typeout{** WARNING: IEEEtran.bst: No hyphenation pattern has been}%
\typeout{** loaded for the language `#1'. Using the pattern for}%
\typeout{** the default language instead.}%
\else
\language=\csname l@#1\endcsname
\fi
#2}}
\providecommand{\BIBdecl}{\relax}
\BIBdecl

\bibitem{7377400}
J.~Petajajarvi, K.~Mikhaylov, A.~Roivainen, T.~Hanninen, and M.~Pettissalo, ``On the coverage of lpwans: range evaluation and channel attenuation model for lora technology,'' in \emph{2015 14th International Conference on ITS Telecommunications (ITST)}, 2015, pp. 55--59.

\bibitem{gupta2024utilizingtransferlearningpretrained}
\BIBentryALTinterwordspacing
H.~P. Gupta and R.~Mishra, ``Utilizing transfer learning and pre-trained models for effective forest fire detection: A case study of uttarakhand,'' 2024. [Online]. Available: \url{https://arxiv.org/abs/2410.06743}
\BIBentrySTDinterwordspacing

\bibitem{1321026}
X.~Wang, J.~Dong, C.~Chin, S.~Hettiarachchi, and D.~Zhang, ``Semantic space: an infrastructure for smart spaces,'' \emph{IEEE Pervasive Computing}, vol.~3, no.~3, pp. 32--39, 2004.

\bibitem{conf/networking/PandeyKGRR24}
\BIBentryALTinterwordspacing
S.~Pandey, P.~Kumari, H.~P. Gupta, D.~Rai, and S.~V. Rao, ``A site-specific lorawan parameters selection approach with multi-loss propagation model.'' in \emph{IFIP Networking}.\hskip 1em plus 0.5em minus 0.4em\relax IEEE, 2024, pp. 350--358. [Online]. Available: \url{http://dblp.uni-trier.de/db/conf/networking/networking2024.html#PandeyKGRR24}
\BIBentrySTDinterwordspacing

\bibitem{8480255}
K.~Mekki, E.~Bajic, F.~Chaxel, and F.~Meyer, ``Overview of cellular lpwan technologies for iot deployment: Sigfox, lorawan, and nb-iot,'' in \emph{2018 IEEE International Conference on Pervasive Computing and Communications Workshops (PerCom Workshops)}, 2018, pp. 197--202.

\bibitem{conf/wcnc/PandeyKG024}
\BIBentryALTinterwordspacing
S.~Pandey, P.~Kumari, H.~P. Gupta, and S.~V. Rao, ``Advancing lorawan network efficiency through dynamic receive window adjustment.'' in \emph{WCNC}.\hskip 1em plus 0.5em minus 0.4em\relax IEEE, 2024, pp. 1--6. [Online]. Available: \url{http://dblp.uni-trier.de/db/conf/wcnc/wcnc2024.html#PandeyKG024}
\BIBentrySTDinterwordspacing

\bibitem{10176016}
D.~Zorbas, D.~Hackett, and B.~O'Flynn, ``On the coexistence of lora and rf power transfer,'' in \emph{2023 IEEE International Instrumentation and Measurement Technology Conference (I2MTC)}, 2023, pp. 1--5.

\bibitem{conf/wcnc/ThangadoraiSGK24}
\BIBentryALTinterwordspacing
K.~K. Thangadorai, K.~M. Sivalingam, H.~P. Gupta, and M.~R. Kanagarathinam, ``Stickyless: An intelligent method for solving sticky client problem in wi-fi networks.'' in \emph{WCNC}.\hskip 1em plus 0.5em minus 0.4em\relax IEEE, 2024, pp. 1--6. [Online]. Available: \url{http://dblp.uni-trier.de/db/conf/wcnc/wcnc2024.html#ThangadoraiSGK24}
\BIBentrySTDinterwordspacing

\bibitem{conf/wcnc/0001G24}
\BIBentryALTinterwordspacing
R.~Mishra and H.~P. Gupta, ``A federated learning approach to minimize communication rounds using noise rectification.'' in \emph{WCNC}.\hskip 1em plus 0.5em minus 0.4em\relax IEEE, 2024, pp. 1--6. [Online]. Available: \url{http://dblp.uni-trier.de/db/conf/wcnc/wcnc2024.html#0001G24}
\BIBentrySTDinterwordspacing

\bibitem{5558084}
L.~Filipponi, A.~Vitaletti, G.~Landi, V.~Memeo, G.~Laura, and P.~Pucci, ``Smart city: An event driven architecture for monitoring public spaces with heterogeneous sensors,'' in \emph{2010 Fourth International Conference on Sensor Technologies and Applications}, 2010, pp. 281--286.

\bibitem{Qin2020}
Y.~Qin, H.~Matsutani, and M.~Kondo, ``A selective model aggregation approach in federated learning for online anomaly detection.''\hskip 1em plus 0.5em minus 0.4em\relax IEEE, 11 2020, pp. 684--691.

\bibitem{Shubyn2023}
B.~Shubyn, D.~Kostrzewa, P.~Grzesik, P.~Benecki, T.~Maksymyuk, V.~Sunderam, J.~H. Syu, J.~C.~W. Lin, and D.~Mrozek, ``Federated learning for improved prediction of failures in autonomous guided vehicles,'' \emph{Journal of Computational Science}, vol.~68, p. 101956, 4 2023.

\bibitem{1589116}
A.~Juels, ``Rfid security and privacy: a research survey,'' \emph{IEEE Journal on Selected Areas in Communications}, vol.~24, no.~2, pp. 381--394, 2006.

\bibitem{journals/tpds/MishraGBD24}
\BIBentryALTinterwordspacing
R.~Mishra, H.~P. Gupta, G.~Banga, and S.~K. Das, ``Fed-rac: Resource-aware clustering for tackling heterogeneity of participants in federated learning.'' \emph{IEEE Trans. Parallel Distributed Syst.}, vol.~35, no.~7, pp. 1207--1220, 07 2024. [Online]. Available: \url{http://dblp.uni-trier.de/db/journals/tpds/tpds35.html#MishraGBD24}
\BIBentrySTDinterwordspacing

\bibitem{7815384}
U.~Raza, P.~Kulkarni, and M.~Sooriyabandara, ``Low power wide area networks: An overview,'' \emph{IEEE Communications Surveys \& Tutorials}, vol.~19, no.~2, pp. 855--873, 2017.

\bibitem{conf/sensys/KumariG023}
\BIBentryALTinterwordspacing
P.~Kumari, H.~P. Gupta, and B.~Sikdar, ``Poster abstract: Efficient knowledge distillation to train lightweight neural network for heterogeneous edge devices.'' in \emph{SenSys}, M.~R. Eskicioglu, P.~Huang, and N.~Patwari, Eds.\hskip 1em plus 0.5em minus 0.4em\relax ACM, 2023, pp. 546--547. [Online]. Available: \url{http://dblp.uni-trier.de/db/conf/sensys/sensys2023.html#KumariG023}
\BIBentrySTDinterwordspacing

\bibitem{7488250}
W.~Shi, J.~Cao, Q.~Zhang, Y.~Li, and L.~Xu, ``Edge computing: Vision and challenges,'' \emph{IEEE Internet of Things Journal}, vol.~3, no.~5, pp. 637--646, 2016.

\bibitem{conf/iccps/0012G023}
\BIBentryALTinterwordspacing
A.~Gupta, H.~P. Gupta, and S.~K. Das, ``Fedar+: A federated learning approach to appliance recognition with mislabeled data in residential environments.'' in \emph{ICCPS}, S.~Mitra, N.~Venkatasubramanian, A.~Dubey, L.~Feng, M.~Ghasemi, and J.~Sprinkle, Eds.\hskip 1em plus 0.5em minus 0.4em\relax ACM, 2023, pp. 78--87. [Online]. Available: \url{http://dblp.uni-trier.de/db/conf/iccps/iccps2023.html#0012G023}
\BIBentrySTDinterwordspacing

\bibitem{9166711}
A.~Gupta, H.~P. Gupta, B.~Biswas, and T.~Dutta, ``An unseen fault classification approach for smart appliances using ongoing multivariate time series,'' \emph{IEEE Transactions on Industrial Informatics}, vol.~17, no.~6, pp. 3731--3738, 2021.

\bibitem{conf/icc/KumariGDB23}
\BIBentryALTinterwordspacing
P.~Kumari, H.~P. Gupta, S.~K. Das, and R.~Bansal, ``Rate-monotonic scheduler for lora-based smart space monitoring system.'' in \emph{ICC}.\hskip 1em plus 0.5em minus 0.4em\relax IEEE, 2023, pp. 2498--2503. [Online]. Available: \url{http://dblp.uni-trier.de/db/conf/icc/icc2023.html#KumariGDB23}
\BIBentrySTDinterwordspacing

\bibitem{journals/csur/MishraG23}
\BIBentryALTinterwordspacing
R.~Mishra and H.~P. Gupta, ``Transforming large-size to lightweight deep neural networks for iot applications.'' \emph{ACM Comput. Surv.}, vol.~55, no.~11, pp. 233:1--233:35, 11 2023. [Online]. Available: \url{http://dblp.uni-trier.de/db/journals/csur/csur55.html#MishraG23}
\BIBentrySTDinterwordspacing

\bibitem{9311219}
P.~Escobedo, M.~Bhattacharjee, F.~Nikbakhtnasrabadi, and R.~Dahiya, ``Smart bandage with wireless strain and temperature sensors and batteryless nfc tag,'' \emph{IEEE Internet of Things Journal}, vol.~8, no.~6, pp. 5093--5100, 2021.

\bibitem{9479778}
A.~Gupta and H.~P. Gupta, ``Yogahelp: Leveraging motion sensors for learning correct execution of yoga with feedback,'' \emph{IEEE Transactions on Artificial Intelligence}, vol.~2, no.~4, pp. 362--371, 2021.

\bibitem{journals/wpc/ChopadeGD23}
\BIBentryALTinterwordspacing
S.~Chopade, H.~P. Gupta, and T.~Dutta, ``Survey on sensors and smart devices for iot enabled intelligent healthcare system.'' \emph{Wirel. Pers. Commun.}, vol. 131, no.~3, pp. 1957--1995, 08 2023. [Online]. Available: \url{http://dblp.uni-trier.de/db/journals/wpc/wpc131.html#ChopadeGD23}
\BIBentrySTDinterwordspacing

\bibitem{conf/infocom/MishraGD22}
\BIBentryALTinterwordspacing
R.~Mishra, H.~P. Gupta, and T.~Dutta, ``Noise-resilient federated learning: Suppressing noisy labels in the local datasets of participants.'' in \emph{INFOCOM Workshops}.\hskip 1em plus 0.5em minus 0.4em\relax IEEE, 2022, pp. 1--2. [Online]. Available: \url{http://dblp.uni-trier.de/db/conf/infocom/infocom2022w.html#MishraGD22}
\BIBentrySTDinterwordspacing

\bibitem{Wang2021}
B.~Wang, F.~Tao, X.~Fang, C.~Liu, Y.~Liu, and T.~Freiheit, ``Smart manufacturing and intelligent manufacturing: A comparative review,'' \emph{Engineering}, vol.~7, pp. 738--757, 6 2021.

\bibitem{conf/wowmom/KumariGDD22}
\BIBentryALTinterwordspacing
P.~Kumari, H.~P. Gupta, T.~Dutta, and S.~K. Das, ``Improving age of information with interference problem in long-range wide area networks.'' in \emph{WoWMoM}.\hskip 1em plus 0.5em minus 0.4em\relax IEEE, 2022, pp. 137--146. [Online]. Available: \url{http://dblp.uni-trier.de/db/conf/wowmom/wowmom2022.html#KumariGDD22}
\BIBentrySTDinterwordspacing

\bibitem{conf/mswim/KumariGM021}
P.~Kumari, H.~P. Gupta, R.~Mishra, and S.~K. Das, ``An energy-efficient smart space system using lora network with deadline and security constraints.'' in \emph{MSWiM}, M.~Aguilar-Igartua, C.~Giannelli, and J.~Zheng, Eds.\hskip 1em plus 0.5em minus 0.4em\relax ACM, 2021, pp. 79--86.

\bibitem{gupta2024lessonslearnedsmartcampus}
\BIBentryALTinterwordspacing
H.~P. Gupta, ``Lessons learned: A smart campus environment using lorawan,'' 2024. [Online]. Available: \url{https://arxiv.org/abs/2410.09927}
\BIBentrySTDinterwordspacing

\bibitem{conf/wowmom/MishraKGSDSP21}
\BIBentryALTinterwordspacing
R.~Mishra, P.~Kumari, H.~P. Gupta, D.~Shrivastava, T.~Dutta, D.~Y. Suh, and M.~J. Piran, ``A game theory-based transportation system using fog computing for passenger assistance.'' in \emph{WOWMOM}.\hskip 1em plus 0.5em minus 0.4em\relax IEEE, 2021, pp. 1--10. [Online]. Available: \url{http://dblp.uni-trier.de/db/conf/wowmom/wowmom2021.html#MishraKGSDSP21}
\BIBentrySTDinterwordspacing

\bibitem{4460126}
J.-S. Lee, Y.-W. Su, and C.-C. Shen, ``A comparative study of wireless protocols: Bluetooth, uwb, zigbee, and wi-fi,'' in \emph{IECON 2007 - 33rd Annual Conference of the IEEE Industrial Electronics Society}, 2007, pp. 46--51.

\bibitem{conf/infocom/KumariGD21}
\BIBentryALTinterwordspacing
P.~Kumari, H.~P. Gupta, and T.~Dutta, ``A lightweight compression-based energy-efficient smart metering system in long-range network.'' in \emph{INFOCOM Workshops}.\hskip 1em plus 0.5em minus 0.4em\relax IEEE, 2021, pp. 1--2. [Online]. Available: \url{http://dblp.uni-trier.de/db/conf/infocom/infocom2021w.html#KumariGD21}
\BIBentrySTDinterwordspacing

\bibitem{conf/icc/KumariGD20}
\BIBentryALTinterwordspacing
------, ``A nodes scheduling approach for effective use of gateway in dense lora networks.'' in \emph{ICC}.\hskip 1em plus 0.5em minus 0.4em\relax IEEE, 2020, pp. 1--6. [Online]. Available: \url{http://dblp.uni-trier.de/db/conf/icc/icc2020.html#KumariGD20}
\BIBentrySTDinterwordspacing

\bibitem{journals/cem/GhoshMGSR22}
\BIBentryALTinterwordspacing
U.~Ghosh, H.~Maziku, H.~P. Gupta, B.~Sikdar, and J.~J. P.~C. Rodrigues, ``Security, trust, and privacy solutions for intelligent internet of vehicular things - part i.'' \emph{IEEE Consumer Electron. Mag.}, vol.~11, no.~6, pp. 39--40, 2022. [Online]. Available: \url{http://dblp.uni-trier.de/db/journals/cem/cem11.html#GhoshMGSR22}
\BIBentrySTDinterwordspacing

\bibitem{9130098}
R.~Mishra, H.~P. Gupta, and T.~Dutta, ``A road health monitoring system using sensors in optimal deep neural network,'' \emph{IEEE Sensors Journal}, vol.~21, no.~14, pp. 15\,527--15\,534, 2021.

\bibitem{9164991}
R.~Mishra, A.~Gupta, H.~P. Gupta, and T.~Dutta, ``A sensors based deep learning model for unseen locomotion mode identification using multiple semantic matrices,'' \emph{IEEE Transactions on Mobile Computing}, vol.~21, no.~3, pp. 799--810, 2022.

\bibitem{10122600}
M.~Jouhari, N.~Saeed, M.-S. Alouini, and E.~M. Amhoud, ``A survey on scalable lorawan for massive iot: Recent advances, potentials, and challenges,'' \emph{IEEE Communications Surveys \& Tutorials}, vol.~25, no.~3, pp. 1841--1876, 2023.

\bibitem{7745306}
M.~B. Yassein, W.~Mardini, and A.~Khalil, ``Smart homes automation using z-wave protocol,'' in \emph{2016 International Conference on Engineering \& MIS (ICEMIS)}, 2016, pp. 1--6.

\bibitem{8170296}
Y.~Li, X.~Cheng, Y.~Cao, D.~Wang, and L.~Yang, ``Smart choice for the smart grid: Narrowband internet of things (nb-iot),'' \emph{IEEE Internet of Things Journal}, vol.~5, no.~3, pp. 1505--1515, 2018.

\bibitem{8502812}
W.~Ayoub, A.~E. Samhat, F.~Nouvel, M.~Mroue, and J.-C. Prévotet, ``Internet of mobile things: Overview of lorawan, dash7, and nb-iot in lpwans standards and supported mobility,'' \emph{IEEE Communications Surveys \& Tutorials}, vol.~21, no.~2, pp. 1561--1581, 2019.

\bibitem{7880946}
M.~Lauridsen, I.~Z. Kovacs, P.~Mogensen, M.~Sorensen, and S.~Holst, ``Coverage and capacity analysis of lte-m and nb-iot in a rural area,'' in \emph{2016 IEEE 84th Vehicular Technology Conference (VTC-Fall)}, 2016, pp. 1--5.

\bibitem{7803607}
O.~Georgiou and U.~Raza, ``Low power wide area network analysis: Can lora scale?'' \emph{IEEE Wireless Communications Letters}, vol.~6, no.~2, pp. 162--165, 2017.

\bibitem{7123563}
A.~Al-Fuqaha, M.~Guizani, M.~Mohammadi, M.~Aledhari, and M.~Ayyash, ``Internet of things: A survey on enabling technologies, protocols, and applications,'' \emph{IEEE Communications Surveys \& Tutorials}, vol.~17, no.~4, pp. 2347--2376, 2015.

\bibitem{7498684}
M.~Chiang and T.~Zhang, ``Fog and iot: An overview of research opportunities,'' \emph{IEEE Internet of Things Journal}, vol.~3, no.~6, pp. 854--864, 2016.

\bibitem{8016573}
Y.~Mao, C.~You, J.~Zhang, K.~Huang, and K.~B. Letaief, ``A survey on mobile edge computing: The communication perspective,'' \emph{IEEE Communications Surveys \& Tutorials}, vol.~19, no.~4, pp. 2322--2358, 2017.

\bibitem{6054047}
C.~Tran and M.~M. Trivedi, ``3-d posture and gesture recognition for interactivity in smart spaces,'' \emph{IEEE Transactions on Industrial Informatics}, vol.~8, no.~1, pp. 178--187, 2012.

\bibitem{8326735}
H.-C. Lee and K.-H. Ke, ``Monitoring of large-area iot sensors using a lora wireless mesh network system: Design and evaluation,'' \emph{IEEE Transactions on Instrumentation and Measurement}, vol.~67, no.~9, pp. 2177--2187, 2018.

\bibitem{7460727}
Y.~Zeng, P.~H. Pathak, and P.~Mohapatra, ``Wiwho: Wifi-based person identification in smart spaces,'' in \emph{2016 15th ACM/IEEE International Conference on Information Processing in Sensor Networks (IPSN)}, 2016, pp. 1--12.

\bibitem{6780609}
S.~Amendola, R.~Lodato, S.~Manzari, C.~Occhiuzzi, and G.~Marrocco, ``Rfid technology for iot-based personal healthcare in smart spaces,'' \emph{IEEE Internet of Things Journal}, vol.~1, no.~2, pp. 144--152, 2014.

\bibitem{Charef2023}
N.~Charef, A.~B. Mnaouer, M.~Aloqaily, O.~Bouachir, and M.~Guizani, ``Artificial intelligence implication on energy sustainability in internet of things: A survey,'' \emph{Information Processing and Management}, vol.~60, p. 103212, 3 2023.

\bibitem{Sikder2023}
M.~N.~K. Sikder and F.~A. Batarseh, ``Outlier detection using ai: a survey,'' \emph{AI Assurance: Towards Trustworthy, Explainable, Safe, and Ethical AI}, pp. 231--291, 1 2023.

\bibitem{Samariya2023}
\BIBentryALTinterwordspacing
D.~Samariya and A.~Thakkar, ``A comprehensive survey of anomaly detection algorithms,'' \emph{Annals of Data Science}, vol.~10, pp. 829--850, 6 2023. [Online]. Available: \url{https://link.springer.com/article/10.1007/s40745-021-00362-9}
\BIBentrySTDinterwordspacing

\bibitem{Kea2023}
\BIBentryALTinterwordspacing
K.~Kea, Y.~Han, and T.~K. Kim, ``Enhancing anomaly detection in distributed power systems using autoencoder-based federated learning,'' \emph{PLOS ONE}, vol.~18, p. e0290337, 8 2023. [Online]. Available: \url{https://journals.plos.org/plosone/article?id=10.1371/journal.pone.0290337}
\BIBentrySTDinterwordspacing

\bibitem{Rodriguez2023}
M.~Rodríguez, D.~P. Tobón, and D.~Múnera, ``Anomaly classification in industrial internet of things: A review,'' \emph{Intelligent Systems with Applications}, vol.~18, p. 200232, 5 2023.

\bibitem{5567086}
D.~Cook, ``Learning setting-generalized activity models for smart spaces,'' \emph{IEEE Intelligent Systems}, vol.~27, no.~1, pp. 32--38, 2012.

\bibitem{Mohammadi2023}
\BIBentryALTinterwordspacing
M.~Mohammadi, R.~Shrestha, S.~Sinaei, S.~A. Salcines, S.~D. Pampliega, S.~Electric, S.~R. Clemente, S.~A.~L. Sanz, A.~Sal-cines, D.~Pampliega, R.~Clemente, and A.~L. Sanz, ``Anomaly detection using lstm-autoencoder in smart grid: A federated learning approach,'' \emph{ACM International Conference Proceeding Series}, pp. 48--54, 8 2023. [Online]. Available: \url{https://dl.acm.org/doi/10.1145/3616131.3616138}
\BIBentrySTDinterwordspacing

\end{thebibliography}
\end{document}